\documentclass[journal=jacsat,manuscript=article]{achemso}

\usepackage[version=3]{mhchem} 

\author{Fabian O. von Rohr}
\affiliation{Department of Quantum Matter Physics, University of Geneva, Quai Ernest-Ansermet 24, CH-1211 Geneva, Switzerland}

\title[An \textsf{achemso} demo]
  {Chemical Principles of Intrinsic Topological Superconductors}

\keywords{Chemical Principles of Materials, Topological Matter, Topological Superconductivity, Quantum Materials}

\begin{document}

\begin{tocentry}

\includegraphics[width=1\textwidth]{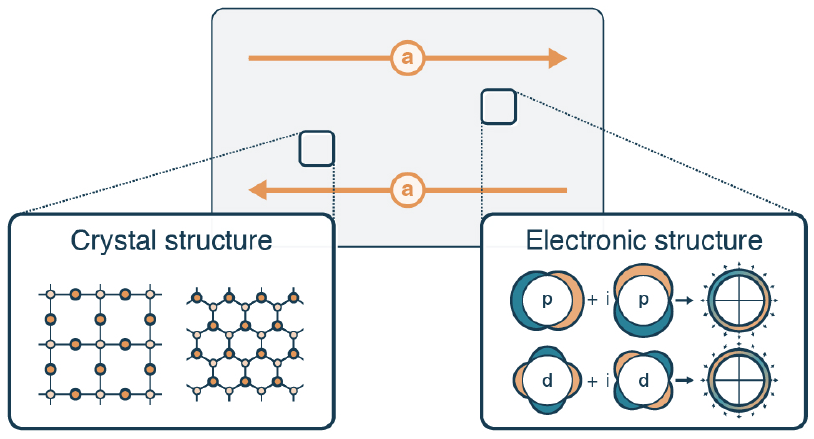}

\end{tocentry}

\begin{abstract}
The concept of topological superconductivity has attracted immense interest in the physics community recently for several reasons: First, topological superconductors represent new phases of matter, topologically distinct from any other known phase of matter. Second, their discovery would present the first realization of Majorana zero modes. Third, intrinsic topological superconductors promise to become an important ingredient for next-generation quantum technologies. There are a handful of candidates to date considered as potential intrinsic topological superconductors. All of these display signs of unconventional, potentially topological superconductivity. However, the results from different experimental methods are inconclusive. One of the major challenges in the field of topological superconductivity has been the scarcity of potential materials, in contrast to the many exciting, theoretical predictions of properties that could be unraveled by such a discovery. Currently, it should be far to say that a material that convincingly displays intrinsic topological superconductivity, and Majorana zero modes, has so far not been discovered. This perspective aims at summarizing the results of the most actively discussed potential topological superconductors for chemists. But, I will also compile the essential physical and chemical design principles of these materials from a chemists' perspective in order to motivate synthetic chemists to join the quest for the discovery of the first intrinsic topological superconductor.
\end{abstract}

\section{Introduction}

Topology is a mathematical concept for the classification of geometric objects. Two objects that can be continuously transformed into one another belong to the same topological class. Topology is in chemistry most commonly used to distinguish between chemical structures. Most prominently to describe supramolecular organic molecules, e.g.,  molecular knots.\cite{frisch1961chemical,fielden2017molecular} The topological analysis of crystal structures is for example used for studying the topological symmetry of complex ionic compounds. A successful example was the proof of the cubic \ce{SiP2O7} not simply being a highly symmetric version of one of the monoclinic polymorphs of \ce{SiP2O7}.\cite{tillmanns1973computer}

Generally, the concept of topology can also be used to describe different electronic phases. In analogy to the real space arrangement of geometric objects, wave functions that are adiabatically connected to each other are topologically identical \cite{Thouless1982}. As a consequence, quantum many-body systems can be topologically trivial or non-trivial, depending on whether or not their wave functions are adiabatically connected \cite{Bernevig2006,Kane2005_2}. The discovery of the quantum Hall effect (QHE) in GaAs/GaAlAs heterostructures by Klaus von Klitzing and co-workers is widely considered to be the very foundation of the classification of matter by topology \cite{von1986}. One of the striking features of the QHE is that by applying a perpendicular magnetic field $B$ on a 2D electron gas (2DEG), there have to appear electronic states localized at the edges of the sample. In a simplified manner: At the edge of the "material" a metallic current is being realized by the external magnetic field, which can be understood as a chiral edge states. This chiral edge state is remarkably robust and so-called topologically protected, as it cannot be simply perturbed. In these heterostructures, various distinguishable topological phases have been identified that do not require symmetry breaking \cite{Bernevig2006,Kane2005}. Commonly, different electronic phases are distinguished by symmetries that they break, these states are, however, distinguished by their different topology. For a considerable amount of time the QHE and some variations of it, were the only realization of topologically non-trivial electronic states in solids. 

In the past few years, especially in the years after 2007, groundbreaking progress has been made in the understanding, prediction, and realization of topologically non-trivial materials. One might even state that there has been a revolution in condensed matter physics as it was realized that many materials, their band-structures and eventually their properties, can be classified and understood in terms of their topology.

Especially, noteworthy is the topological band theory by Kosterlitz and Thouless, the Kane-Mele model, the prediction, and realization of the 2D quantum hall effect, the prediction, and realization of 3D topological insulators, as-well as Dirac and Weyl semimetals. This is a highly subjective selection, and there are many more important manuscript that appeared on the topic of topological materials in the last $\approx$15 years. It can be noted, that with the prediction and discovery of topological insulators and topological semimetals, topologically non-trivial materials have attracted immense interest in the physics community over the past decade. There are several extensive reviews on the chemistry, crystallographic structure, and composition of topological insulators and topological semimetals available now that the interested reader may be directed to (see references \citenum{Hasan2010,Burkov2011,muchler2012topological,Ando2013,cava2013crystal,Schoop2018,kumar2020topological}). 

The field of research of topologically non-trivial materials remains fast-developing. It is driven by exciting predictions of new physical phenomena and novel types of quantum matter that can arise in topological materials, due to the unusual nature of the charge carriers in these materials. One of the most exciting developments is the postulation of the possibility to discover a topological superconductor. Such a superconductor is topologically distinct from a conventional superconductor, as a result it must have so-called Majorana zero-modes on its surface, edges and/or in its vortex cores. These Majoranas are of great interest for the potential realization of topological quantum computers, which are expected to be fault-tolerant in contrary to the currently developed quantum computers \cite{kitaev2003fault,Nayak2008}. It is reasonable to assert that there has been no discovery of a material that demonstrates intrinsic topological superconductivity and Majorana zero modes in a convincingly definitive manner, yet. As I will argue in the following, this remains likely to a large extent a solid-state chemistry challenge, as the development and discovery of new materials appears to be an indispensable necessity for the realization of intrinsic topological superconductivity.

\begin{figure}
	\centering
	\includegraphics[width=0.4\textwidth]{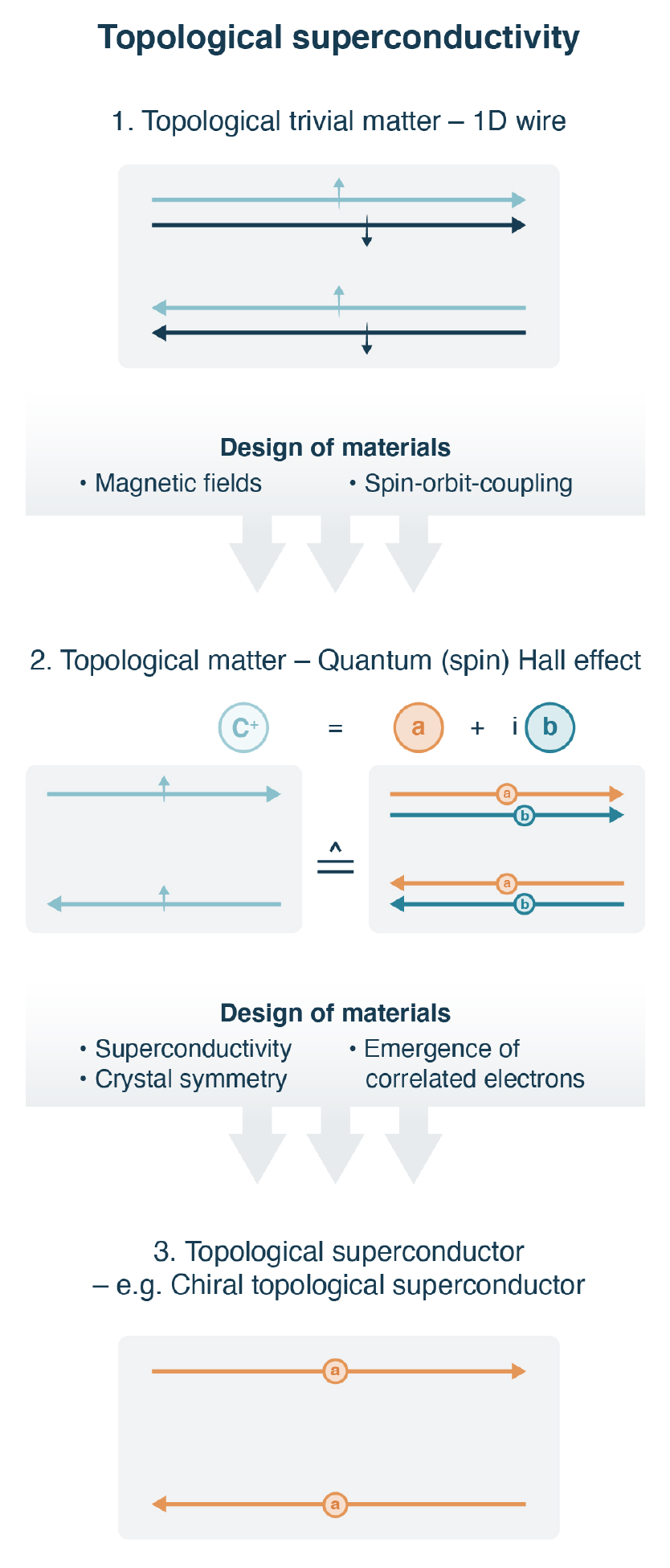}
	\captionof{figure}{From topological materials to topological superconductors, a question of appropriate materials discovery and/or design.}
	\label{fig:1}
\end{figure}

\section{The Basics of Topological Superconductivity}

In this perspective, we will concentrate on so-called intrinsic topological superconductors. Topological classifications are possible in principle for various quantum many-body systems that have a gap in the energy-band spectra. Therefore, also a systematic topological classification of superconductors is possible and theoretically straightforward, while the real-world realization of topological superconductors has been proven to be challenging and remains elusive.

The symmetry and the dimensionality of the electronic energy gap of a superconductor in the unpaired particle spectra is in this case used to define the corresponding topological invariants. In general, topological superconductors are adiabatically distinct from the Bose--Einstein condensate of Cooper pairs, which forms a topologically trivial superconducting state \cite{Bernevig2013}.

The quest for finding topological superconductors has really been somewhat emerging after the prediction and realization of 3D topological insulators \cite{Sato2017,Li2019}. While it is fair to say that much of the current work is strongly affected and inspired by the proposal by \textit{Liang Fu} and \textit{Charles Kane} in 2008\cite{Fu2008}. However, theoretical work on this topic actually dates back to Volovik, who discussed non-trivial topology in superfluid \ce{^3He} already in 2003 \cite{Volovik2003}, though the vocabulary to describe these phenomena and the broader context had been somewhat lacking then. Due to the historical chronology of events, materials with non-trivial electronic properties have been investigated widely for the possible emergence of topological superconductivity. The relationship and the interconnection of topological materials and topological superconductors are graphically summarized in figure \ref{fig:1}. The importance of materials design and/or materials discovery is clearly highlighted in this representation. Only the design of materials allowed for the realization of topological materials from topologically trivial matter in the first place. Similarly, it is once more a design and/or discovery question now to realize by analogy a topological superconductor.

The strong spin-orbit coupling present in many topological materials, and/or the strong restrictions electrons might experience in topologically non-trivial bands at the Fermi-surface, might be beneficial to the occurrence of topological superconductivity. However, the mere (co-)existence of non-trivial topology and superconductivity in a material are not sufficient conditions for the occurrence of topological superconductivity. Nevertheless, the interplay of topological states and collective electronic states in general has so far only been insufficiently investigated. These coexistence and their interactions are certainly a topic of great current research. However, for the discovery of an intrinsic topological superconductor, further conditions have to be met, which we will be discussing in the following. 

There are three principle approaches through, which we can realize an intrinsic topological superconductor, this is either by a special case of a chiral superconductor, by a helical superconductor, or by a weak/higher-order topological superconductor  (these cases will be discussed below). A fourth way to obtain a topological superconductor, is strongly motivated by an ingenious proposal by \textit{Liang Fu} and \textit{Charles Kane}, where they proposed Majorana fermions at the surface of a topological insulator via a proximity effect.\cite{Fu2008} The authors of this seminal work showed that when a superconductor is coupled to a topological insulator, a topologically non-trivial pairing state appears at the surface of the contact as a projection. To date, there is some, albeit inconclusive and controversially debated, evidence that in these and related artificial heterostructures, Majorana bound states occur \cite{Alicea2012,Mourik2012,Hart2014,Wiedenmann2016,Jack2019,Lutchyn2010,Peng2015,Ren2019}. However, these are confined to the respective nanostructure and are a result of the artificially created structure. This fourth approach to a topological superconductor can be achieved on the surface of a topological material. This type of quantum materials "engineering" by nano-fabrication is -- along with related techniques -- currently, the most widely chosen approach to aim for the realization of topological superconductivity by physics groups around the world. This approach will, however, never result in an intrinsic topological superconductor, but at best in nano-confined domains, where a Majorana zero-mode might be observed. Hence, this approach is not the focus of the perspective at hand. The interested reader may consult the following publications for further information on this topic.\cite{mourik2012signatures,frolov2020topological,jack2019observation} And, any interested reader should also be referred to the important considerable controversy around the topic of Majoranas in some semiconductor nanostructures, where also several manuscripts had to be retracted, for more details see reference \citenum{frolov2021quantum}. Here, we will not discuss these quantum material engineering approaches. But, the focus of this perspective is the current state for the realization of an intrinsic topological superconductor, i.e., superconductors that because of the topologically distinct nature of their electronic structure, need to have Majorana zero-modes in their vortex cores and with Majorana dispersive states on their surface and/or edges.

\subsection*{The Starting Point: Superfluidity in \ce{^3He} and Analogous Superconductivity}

Superfluidity and superconductivity are strongly related in that they are both examples of macroscopic quantum phenomena, where the behavior of large numbers of particles is governed by quantum mechanics. They both involve the condensation of bosons and the absence of friction, they differ in that superconductivity involves the flow of electric current without resistance, while superfluidity involves the flow of a fluid without viscosity. Hence, the superfluidity of \ce{^3He} is an excellent starting point to approach topological superconductivity. 

A superfluid is a fluid with zero viscosity that can flow without losing energy to friction. It typically occurs at very low temperatures, near absolute zero, and is a characteristic of certain quantum fluids. A superfluid can exhibit unusual properties, such as the ability to flow up the sides of a container and even penetrate solid barriers. \ce{^4He} is the most famous superfluid. Specifically, it undergoes a phase transition to a Bose–Einstein condensate at  $T_{\rm c} =$  2.17 K resulting in a superfluid state.\cite{tilley2019superfluidity} 

Superfluidity has also been observed in \ce{^3He}, however, at a much lower temperature of $T_{\rm c} =$ 2.49 mK.\cite{osheroff1972evidence} \ce{^4He}, with its two protons, two neutrons, and two electrons, has a high symmetry and a total nuclear spin of 0, which means it belongs to the class of particles known as bosons. In contrast, \ce{^3He} has a spin of 1/2, which means it is a fermion. Due to this difference in spin, \ce{^3He} has a much higher zero-point energy than \ce{^4He}. This is because the Pauli principle requires all \ce{^3He} atoms to be in different states, while at low temperatures, any number of \ce{^4He} atoms can occupy the same ground state. \ce{^3He} superfluidity results from two \ce{^3He} atoms that form a Cooper pair, while in the case of \ce{^4He}, this intermediate step is not necessary, and it becomes superfluid directly. In the case of \ce{^3He}, several symmetries are broken in its superfluid state, including spin- and orbital-rotation symmetries, mirror symmetry, and time-reversal symmetry. Additionally, \ce{^3He} also breaks U(1) symmetry, which is a characteristic feature of all superconductors and superfluids. Many possible ground states exist for \ce{^3He} in which states with orbital angular momentum equal to $\hbar$ are spin-triplet states with $\hbar$ spin angular momentum. Hence, \ce{^3He} is a $p$-wave superfluid, with spin $S$ = 1 and the angular momentum $L$ = 1.

In zero magnetic field, there are two distinct superfluid phases of \ce{^3He}, the so-called A-phase and the B-phase. The B-phase is the low-temperature, low-pressure phase which has an isotropic energy gap. The A-phase is the higher temperature, higher pressure phase that is further stabilized by a magnetic field and has two point nodes in its gap structure. The presence of two phases is a clear indication that \ce{^3He} is an unconventional superfluid, since the presence of two phases requires an additional symmetry, apart from gauge symmetry, to be broken. The ground state corresponds to total angular momentum zero, $J = S + L = 0$ (vector addition). The superfluid phases of helium have been studied extensively, and the interested reader is directed towards some of the excellent reviews on the topic for further details.\cite{leggett1999superfluidity,khalatnikov2018introduction,schmitt2015introduction} 

The existence of the superfluid A-phase and B-phase of \ce{^3He} are to a large extent motivation and also precedent for the realization of chiral, helical, and eventually topological superconductors.\cite{Read2000,Kitaev2001} The A-phase would in this regard be analogous to the chiral superconductivity first proposed for \ce{Sr2RuO4}.\cite{rice1995} This phase is time-reversal symmetry breaking. The B-phase, however, would be analogous to a 3D $p$-wave helical superconductor with a winding number. This phase would not be time-reversal symmetry breaking. While the helical $p$-wave would be automatically a topological superconductor, the chiral superconductor would only be topologically non-trivial under restrained conditions, e.g., if the superconductivity was very two-dimensional and henceforth the sheets of a $p+ip$ were stacked on one another with no or only very little coupling. These are, therefore, the two main ways to discover an intrinsic topological superconductor. Hence, we will focus on discussing superconductors that may either be analogous to the superfluidity in the A-phase or the B-phase of \ce{^3He}.

\subsection*{How can a superconductor be topologically non-trivial?}

As mentioned above, a topological superconductor is adiabatically distinct from the Bose--Einstein condensate of Cooper pairs, which forms a topologically trivial superconducting state. How can this be possible in a material?

In the superconducting state, there is an electronic energy gap that opens in the unpaired particle spectra, as shown in figure \ref{fig:2}(a). As one can intuitively see in this representation, a superconductor has an energy spectrum that is analogous to the one in an insulator and/or semiconductor.\cite{Tinkham2004} In an insulator/semiconductor, the energy gap is known as the band gap; it represents the amount of energy required to excite an electron from the valence band to the conduction band. In a superconductor, the energy gap is known as the superconducting gap. It represents the amount of energy required to break a Cooper pair apart and return the material to its normal, non-superconducting state. The energy gap of a superconductor is directly related to the particle-hole symmetry, as it is the energy required to break the symmetry between particles and holes. Similarly to the case of the insulator/semiconductor, the superconductor can also be described by a wave function in a mean-field approximation, i.e. the Bogoliubov-deGennes equations.\cite{Tinkham2004} A superconductor can create and annihilate pairs of electrons through the process of breaking apart and forming Cooper pairs. The Hamiltonian of this is written in the following form:

\begin{equation}
H = \sum_{k} (c^\dagger_{k\uparrow}, c_{-k\downarrow})
\begin{pmatrix}
\epsilon(k) & \Delta(k) \\
\Delta(k) & -\epsilon(k)
\end{pmatrix}
\begin{pmatrix}
c_{k\uparrow} \\
c^\dagger_{-k\downarrow}
\end{pmatrix}
\end{equation}

The $c\dagger$ and $c$ are the annihilation and creation operators of the electrons and $\Delta$ being the order parameter. What this -- in a simplified manner -- means, is that also the superconducting state can be represented by a band structure with so-called Bogoliubov quasiparticles. Hence, the excitations of a superconductor are Bogoliubov-De Gennes particles. These form bands, with a gap. And just like “regular” band-structures they can be topological or not, i.e., carry a nontrivial topological index like a Chern number. That, in turn, leads to topologically protected edge states.

The Bogoliubov-deGennes equation has a particle hole symmetry which can be understood in such a fashion that it must be equivalent to either have a hole or an electron in the condensate of Cooper pairs, so having an electron or a hole must be the same. This would actually already be a Majorana state in some sense. However, in a conventional superconductor, the pairing of the Cooper pair is between an up and a down spin electron. Hence, the electron and the holes have opposite spins and are because of that distinct from each other. In order to realize a superconductor this way in which the hole and the electron are not distinct from each other, one needs a spinless superconductor, as we will discuss further in the following sections. 

\subsection*{The Hallmark Feature of TSC: Majorana Zero-modes}

Majorana fermions are a type of particle that have been the subject of intense research in the field of theoretical physics. The concept of Majorana fermions was first proposed by the Italian physicist Ettore Majorana in 1937\cite{Majorana1937}. There are many mysteries around Ettore Majorana, as he disappeared under unresolved circumstances in 1938, and it is unclear what had happened to him\cite{magueijo2010brilliant}. Essentially, Ettore Majorana showed that there exist real solutions to the Dirac equation, which suggests the existence of the so-called Majorana fermion. This particle would possess a half-integer spin, and it would be its own antiparticle. This is in contrast to most other particles, which have distinct antiparticles with opposite charge. Until today, no such Majorana fermions has been found in particle physics, though the Neutrino is still potentially considered to be a Majorana fermion.\cite{bilenky1980oscillations} 

However, as described above such Majoranas, specifically Majorana zero-modes, i.e. quasi-particle, which are their own anti-quasi-particle, can theoretically be realized in a condensed matter system, e.g. in a topological superconductor. As we will see below, these can be either realized in a topological chiral superconductor with broken time-reversal symmetry, or in a helical superconductor.\cite{schnyder2008classification} In both cases, the edge states occur either with a definite chirality for the chiral superconductor or with a helical pair of edge states with counter-propagating spin states in the case of the helical superconductor. This is visualized in figure \ref{fig:2}(c)\&(d). 

Commonly, in condensed matter systems, Majoranas are denoted as zero-modes, and not as fermions. A Majorana zero-mode is actually a type of fermion. It is, however, called a zero-mode because it is localized at a specific point or region and because it is at zero-energy. The reason that a Majorana zero-mode is often referred to as "non-fermionic" is because it has unique properties that distinguish it from ordinary fermions.

Majorana zero-modes are so-called non-Abelian anyons.\cite{van2012coulomb} Hence, they do not follow Abelian statistics, which means that when two Majorana zero-modes are exchanged, the resulting state of the system is not simply a product of the individual states.\cite{beenakker2020search,stern2010non} What non-Abelian means is that the order in which one does things matters. For example, if someone has 3 Majoranas, then moving 1 around 2 and then 2 around 3 leads to a different state than moving 2 around 3 and then 1 around 2. For further details on this, the interested reader is referred to reference [\citenum{alicea2011non}].

This is in contrast to ordinary fermions, where exchanging two fermions simply leads to a sign change in the wave function. This property that leads to their non-Abelian braiding properties is in fact what makes them such a promising platform for so-called fault-tolerant topological quantum computing.\cite{kitaev2003fault,witzel2006quantum} 

The way anyons exchange can be explained by how their world lines interlace in a space-time diagram.\cite{Nayak2008} This is commonly referred to as "braiding," which is similar to the zigzag pattern of interlaced strands of wires or hairs.\cite{teo2010majorana} Topologically unique braids, which cannot be transformed into each other without cutting the world lines, correspond to different unitary matrices that can serve as fundamental components for a quantum computation. Hence, the quantum information is stored in states with multiple quasi-particles, which have a topological degeneracy. The unitary gate operations that are necessary for quantum computation are carried out by braiding quasi-particles and then measuring the multi quasi-particle states. The fault tolerance of a topological quantum computer arises from the non-local encoding of the quasi-particle states, which makes them immune to errors caused by local perturbations.\cite{Nayak2008} This concept of braiding and topological degeneracy is crucial for quantum computers because it provides a robust and fault-tolerant method for processing quantum information. By harnessing the topological properties of anyons, quantum computers can perform complex calculations with high precision, even in the presence of noise and errors, making them more practical for real-world applications.\cite{Nayak2008}

\subsection*{Topological Superconductivity in 1D}

\begin{figure}
	\centering
	\includegraphics[width=0.7\textwidth]{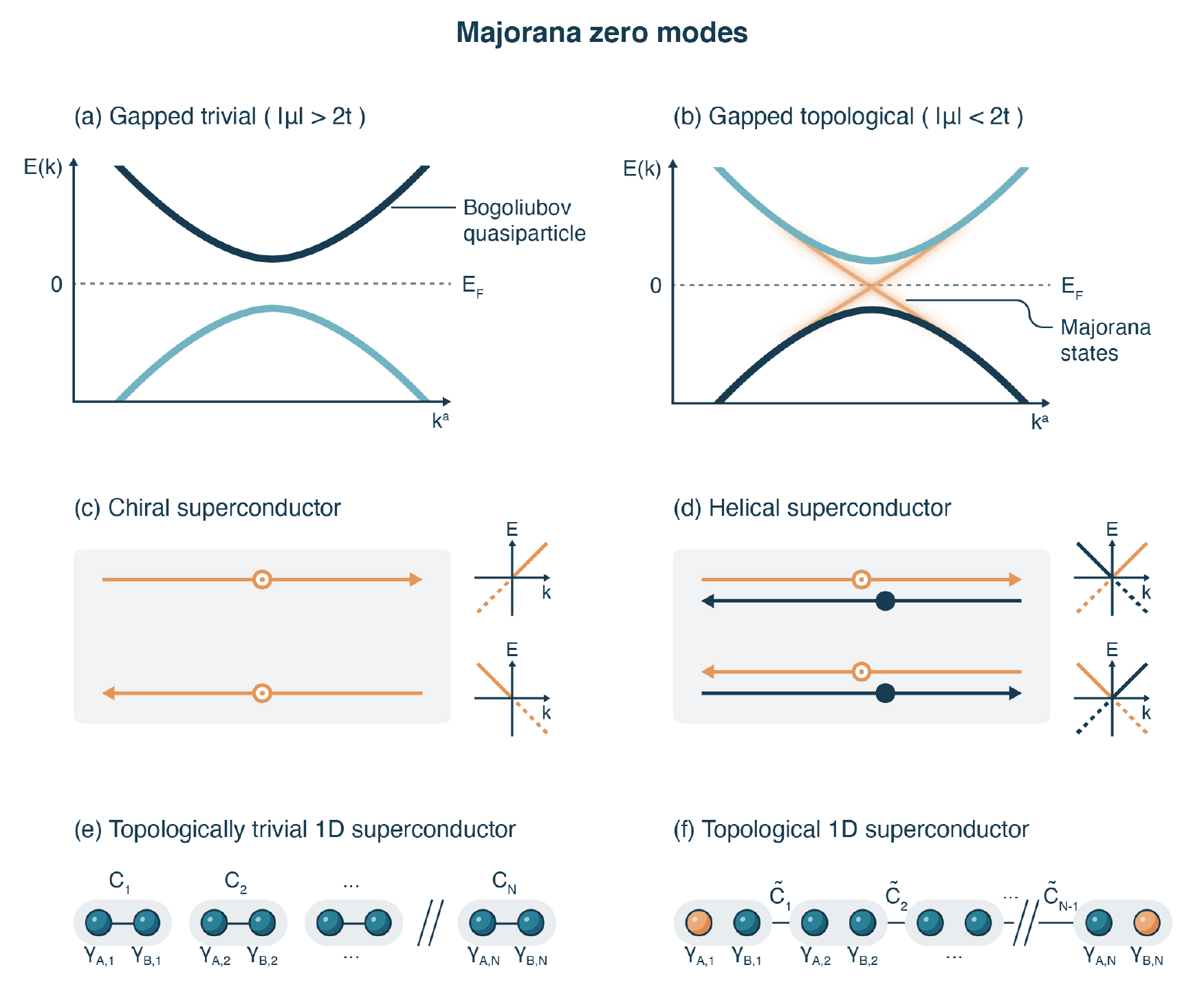}
	\caption{Essentials of topological superconductors. (a) Schematic representation of a Bogoliubov-deGennes bands (a) in a conventional superconductor in case that the chemical potential is smaller than $2t$ and (b) in a topologically non-trivial superconductor with inverted bands and the chemical potential larger than $2t$. (c) Schematic representation of a chiral superconductor with broken time-reversal symmetry, and (d) for a helical superconductor with invariant time-reversal symmetry, both in 2D. In both cases, Majorana states occur either with a definite chirality for the chiral superconductor or with a helical pair of edge states with counter-propagating spin states in the case of the helical superconductor. Figure (c)\&(d) are adapted from reference \cite{Qi2009}. One-dimensional chain of spinless fermions once for (e) the topologically trivial phase, where the Majoranas on each site are coupled so that they form a standard fermion. And, for (f) the topologically non-trivial phase, where the Majoranas on each site are decoupled while Majoranas on adjacent sites become coupled. This leaves two unpaired Majoranas at the ends of the chain.}
	\label{fig:2}
\end{figure}

Understanding and/or visualizing how a superconductor can be topologically distinct from a conventional superconductor is not trivial in 3D. However, when looking at a model in 1D, one can intuitively grasp, how this might be realized. For this, we look at a famous toy model in which Majorana zero-modes can result at the end of the chain superconductor. The end of a 1D chain is equivalent to the surface of a 3D topological superconductor or to the edges of a higher-order topological superconductor in 3D. In a one-dimensional tight-binding representation, a conventional $s$-wave superconductor would just look as shown in figure \ref{fig:2}(a). Where in the bulk there is an excitation gap of 2 $\Delta_0$ in the single particle spectra, due to the formation of Cooper pairs. At the boundary, the gap now closes and goes to zero. The energy gap at the boundary decreases. Superconductors have the intrinsic property that the energy levels always have to be symmetrically arranged around the middle of the gap, due to the aforementioned particle-hole symmetry. As a consequence, there has to be always, for every energy level above zero energy, also an energy level below zero energy. Thus, creating a quasi-particle in state $E$ is equivalent to removing one from state $-E$. One can express this in second quantization by this relation between creation operators and annihilation operators, as

\begin{equation}
\gamma^\dagger(E) = \gamma(-E).
\end{equation}

This is the well-known result from the Majorana equation, representing a particle that is its own anti-particle. This means if any two of these particles meet, they will annihilate each other. If one can realize a state at zero energy, one has a Majorana zero-mode. These Majorana zero-modes are quasi-particles that are their own anti-quasi-particle.\cite{Oreg2010,Sato2017} Unlike in particle physics, particles obey the laws of a Majorana fermions if the creation operator $\gamma^+$ of a quasiparticle, which in this case is a superposition of electron and hole excitations, is identical to the annihilation operator $\gamma$ \cite{Oreg2010,Sato2017}. In the Read--Green model, the Bogoliubov quasiparticles, which correspond to a neutral excitation in a superconductor, become in the bulk material dispersive Majorana fermions, which can be bound to a defect at zero energy in the vortex core. This in turn leads to Majorana zero modes.

However, in the case of a conventional superconductor, it is not possible to have such a state at zero energy, due to zero-point motion. Zero-point motion refers to the quantum mechanical property of particles to retain some residual vibration even at absolute zero temperature. This is because, according to the Heisenberg uncertainty principle, it is not possible to precisely determine both the position and momentum of a particle, so even at 0 Kelvin, particles still possess a non-zero amount of energy.

Such a state at zero energy can, however, be realized in a topological superconductor. The simplest model in which such a state can be designed is in a $p$-wave superconductor in a 1D chain. This possibility was already introduced in 2001 by \textit{Alexei Kitaev}\cite{Kitaev2001,qi2011topological}. Such a system can be written as a 1D tight-binding chain with $p$-wave superconducting pairing, in such a form:

\begin{equation}
H=-\mu \sum_x c_x^{\dagger} c_x-\frac{1}{2} \sum_x\left(t c_x^{\dagger} c_{x+1}+\Delta c_x^{\dagger} c_{x+1}^{\dagger}+\text { H.c. }\right)
\label{eq:kitaev}
\end{equation}

This Hamiltonian is very famous among physicists working on topological superconductors. In essence, it consists in the first part of a chemical potential $\mu$ with an onsite energy, the hopping energy $t$, ${\rm c_{x}}$ is the electron annihilation operator, and this term $\Delta c_x^{\dagger} c_{x+1}^{\dagger}$, which is the superconducting pairing of electrons on neighboring sites. This last expression is commonly referred to as the $p$-wave pairing term, these are spinless fermions. H.c. means hermitian conjugate. 

One can now show that two topologically distinct states are possible in this scenario with this Hamiltonian. These two topologically distinct states depend on the size of the chemical potential $\mu$. The superconducting phase can be topologically trivial in the case that the chemical potential is larger than $2t$, i.e. $\left\vert \mu \right\vert > 2t$. More interestingly, the superconducting phase can be topologically non-trivial with the chemical potential being smaller than $2t$, i.e. $\left\vert \mu \right\vert < 2t$. Both of these superconducting phases are gapped, though the Bogoliubov-deGennes bands are inverted, which is analogously to the inverted bands for a topological superconductor. This is graphically represented in figure \ref{fig:2}(b). As a result of the inversion of these bands the superconductor is topological, and as a result, there has to be a surface and/or edge that connects the topologically non-trivial material to the topological trivial world and/or vacuum. This results in the case of a topological superconductor in a Majorana surface and/or edge state.\cite{bernevig2013topological}

These Majorana states on the surface or at the edges of a topological superconductor can be understood in analogy to the metallic surface state of a topological insulator. Basically, the material is topologically distinct from vacuum, now at the surface there has to be a transition between the two different topologies, which results in a special electronic surface state for the topological insulator, and in Majorana zero-modes for the topological superconductor. A real-world analogy to picture that is, when one wants to build a bridge between the UK and mainland Europe, then the lanes have to cross somewhere in order to change the handedness of the road. This bridge with its properties are a simplified visualization of the surface states of a topological insulator and the Majoranas of a topological superconductor, respectively. In this sense, these special states are an intrinsic property of the material that will always be there. So in a sense we end up at the "chicken or the egg" question.

How can this topological superconductor be understood? Any electron can be written as a combination of two Majoranas. Hence, it is also possible to rewrite the Kitaev model in equation \ref{eq:kitaev} in terms of Majorana operators, which are

\begin{equation}
\begin{aligned}
& c_x=\frac{e^{-i \phi / 2}}{2}\left(\gamma_{B, x}+i \gamma_{A, x}\right) \\
& c_x^{\dagger}=\frac{e^{+i \phi / 2}}{2}\left(\gamma_{B, x}-i \gamma_{A, x}\right).
\end{aligned}
\end{equation}

Resulting in the following Hamiltonian, which is essentially the same as the Hamiltonian in equation \ref{eq:kitaev}:

\begin{equation}
H=-\frac{\mu}{2} \sum_x^N\left(1+i \gamma_{B, x} \gamma_{A, x}\right)-\frac{i}{4} \sum_x^{N-1}(\Delta+t) \gamma_{B, x} \gamma_{A, x+1}+(\Delta-t) \gamma_{A, x} \gamma_{B, x+1}
\end{equation}

One can now show that these two cases -- that we discussed above -- result in two different representations of these Majorana operators in this 1D chain superconductor. In the trivially gapped phase, the coupling of the Majoranas is simply between neighbors, see in figure \ref{fig:2}(e). However, in the topologically non-trivial state, the coupling between the Majoranas is so that there is one Majorana "left" at each end of the 1D chain, as represented in figure \ref{fig:2}(f). Hence, in the topological state, there appear Majorana zero-modes at the edges of the wire. Similarly to this 1D chain case, one can show that there are Majoranas at the edges or the surfaces for any topologically non-trivial superconductor. For further reading on this topic and for the extension of these concepts into 2D and 3D the interested reader is directed towards the following references \citenum{bernevig2013topological,qi2011topological,sato2017topological,leijnse2012introduction}

\begin{figure}
	\centering
	\includegraphics[width=0.5\textwidth]{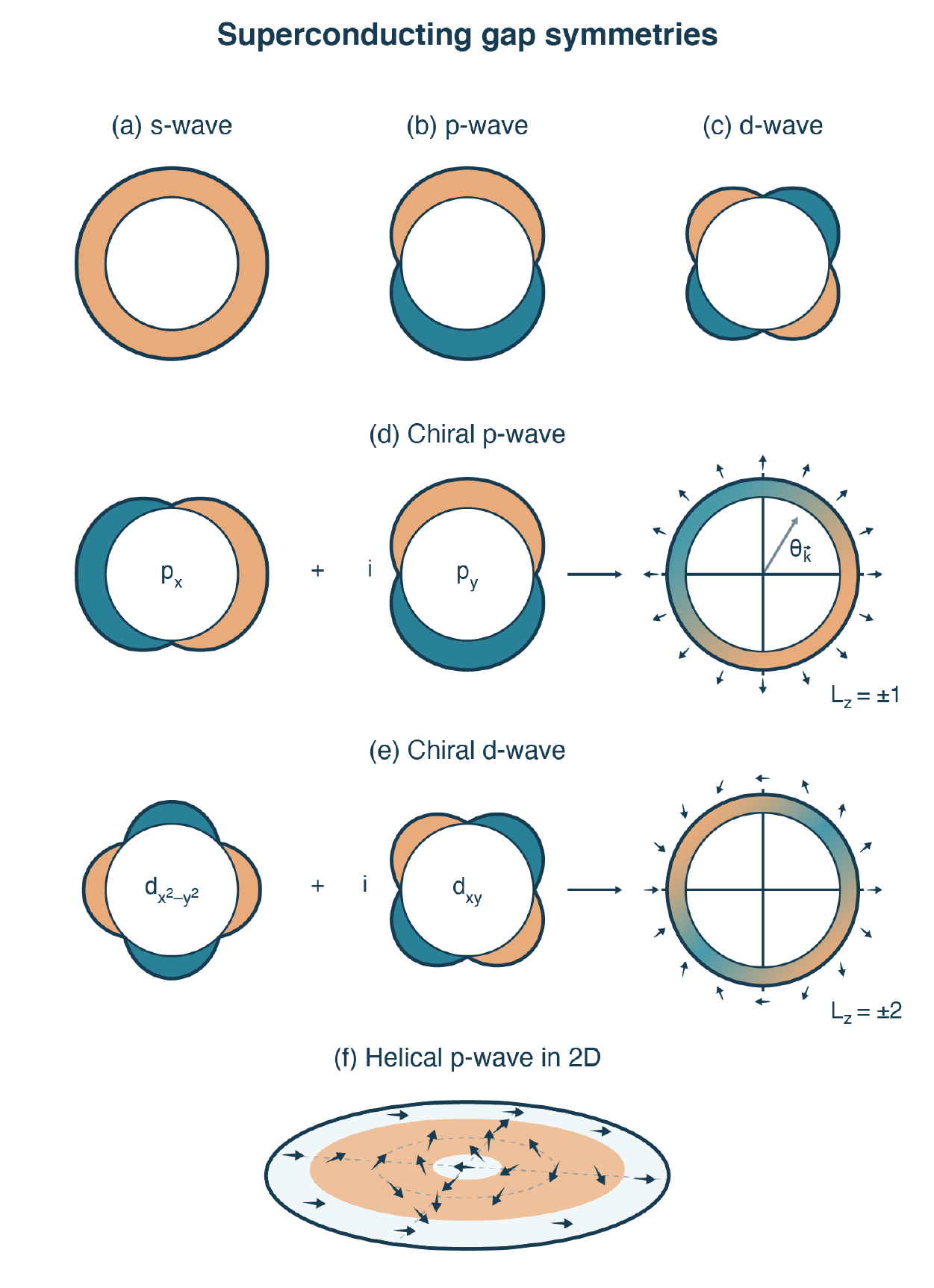}
	\captionof{figure}{Schematic representation of (a) a chiral superconductor with broken time-reversal symmetry and (b) a helical superconductor with invariant time-reversal symmetry. In both cases edge states occur either with a definite chirality for the chiral superconductor or with a helical pair of edge states with counter-propagating spin states in the case of the helical superconductor. The dashed lines show how these edge states of the superconductors are Majorana zero modes, making the $E < 0$ part of the quasiparticle spectra redundant. Figure adapted from reference \cite{Qi2009}. (c) Degenerate $p_x$ and a $p_y$ symmetric gap and (d) a $d_{x^2-y2}$ and a $d_{xy}$ symmetric gap, leading to an overall node-less superconducting gap, and Cooper pairs with an angular momentum.}
	\label{fig:3}
\end{figure}

\section{The Road towards Topological Superconductivity}

Superconductors commonly exhibit $s$-wave symmetry with spin-singlet pairing. That is, the symmetry of the superconducting gap is node-less. The Cooper pairs of these superconductors usually form as electron--electron pairs mediated by phonons. These are referred to as conventional or BCS-type superconductors, as they are described by the Bardeen--Schrieffer--Cooper (BCS) theory of superconductivity \cite{Bardeen1957,Tinkham2004}. All of these superconductors are topologically trivial because they exhibit a smooth crossover from the weak-coupling BCS limit to the strong-coupling Bose--Einstein condensate (BEC) limit without undergoing a gap-closing phase transition \cite{Qi2009}. For the discovery of topological superconductivity, we therefore have to search for superconductors with so-called unconventional pairing states, or for particular combinations of electronic surface states with $s$-wave Cooper pairs.

The total spin $S$ of a Cooper pair can either be spin singlet $S = 0$, or spin triplet $S = 1$.\cite{Tinkham2004} Therefore, odd parity of the wave function of a Cooper pair requires even parity of the orbital wave function $L$ = 0, 2, 4\ldots, satisfying opposite parity of the total wave function. Moreover, $L$ = 0 or 2 corresponds to $s$-wave or $d$-wave pairing, respectively. In spin-triplet states $S = 1$, the orbital wave function has to be odd parity, hence, $L$ = 1 ($p$-wave) or 3 ($f$-wave). The resulting gaps for a circular Fermi surface are shown in figure \ref{fig:3}(a)-(c).

The occurrence of $p$-wave or $d$-wave superconductivity is rare, but both have been observed in real materials. Especially, $d$-wave pairing is widely considered to be proven to exist, as it has been observed with a wide range of methodologies, especially in the cuprate superconductors.\cite{tsuei2000pairing,wollman1995evidence} Examples for candidates for a $d$-wave symmetric superconducting gap are the cuprates or some of the iron-based superconductors \cite{Kirtley1995,Tsuei1997,Guguchia2015,Mazin2010,stock2008spin}. It should be noted that most iron-based superconductors are believed to be unconventional superconductors, but with a spin-singlet sign-changing $s$-wave, usually referred to as s$^{+-}$-wave superconducting gaps on multi-bands.\cite{bang2017superconducting,Guguchia2015}

The existence of $p$-wave superconductivity had been proposed already 60 years ago.\cite{balian1963superconductivity} Since then, theoretical frameworks that govern it have been well established. Detecting materials that demonstrate $p$-wave superconductivity has, however, proven difficult so far. Sensitive probes for $p$-wave superconductivity must interact with either the odd parity or the spin component of the pairing. While the majority of experimental research has focused on the latter. One of the few materials in which $p$-wave superconductivity was thought to exist for two decades was \ce{Sr2RuO4}\cite{rice1995}, but recent experiments have disproven the existence of a $p$-wave order parameter in this material.\cite{pustogow2019constraints} Nowadays, promising candidates for a $p$-wave symmetric superconducting gap are, for example, \ce{UPt3}, \ce{UTe2}, and \ce{CeRh2As2}.\cite{Ran2019,Janoschek2019,Joynt2002,Jiao2019,Norman2011,Tsutsumi2012,khim2021field} 

Gap symmetries other than $s$-wave, $p$-wave, or $d$-wave have not yet been discovered experimentally. The presence of a $p$-wave or $d$-wave symmetry gap (that is, unconventional superconductivity) is therefore a most likely prerequisite for finding intrinsic topological superconductivity.

\section{Types of Intrinsic Topological Superconductors}

As aforementioned, for this perspective, we concentrate on bulk crystals that exhibit so-called intrinsic topological superconductivity. In this case, the topological superconductivity arises as a collective emergent quantum ground state. There are two ways to achieve this: by finding either a special case of a chiral superconductor or finding a helical superconductor. For a complete classification of all possible topological phases in three dimensions -- also including more exotic possibilities than the ones discussed here -- the interested reader is referred to reference \citenum{Schnyder2013}.

\subsubsection*{Chiral Topological Superconductors}

The chiral superconductor that is topologically non-trivial is the analogue to a weak topological insulator, and can be understood as a stacked series of layers of 2D chiral superconductors \cite{Qin2019}. In chiral superconductors feature pairing gaps that wind in phase around the Fermi surface in multiples of 2$\pi$, breaking the time reversal symmetry breaking. The simplest and most widely discussed example is a $k_x + i k_y$ gap function which precesses $\pm 2\pi$ as $\vec{k}$ follows a closed path enclosing the $k_z$ axis.\cite{kallin2016chiral} For a chiral superconductor, symmetry is a highly relevant ingredient, especially the symmetry of the crystal structure. In order to obtain a topological superconductor from a chiral superconductor, one needs to find a state where, for example, a $p_x$ and a $p_y$ symmetric superconducting gap or a $d_{x^2-y^2}$ and a $d_{xy}$ symmetric superconducting gap are degenerate or nearly degenerate \cite{Kallin2016,Qin2019,Kiesel2013}. This can, for example, be realized when a $p$-wave symmetric gap occurs on a square crystallographic lattice, or when a $p$-wave symmetric or a $d$-wave symmetric gap occurs on a hexagonal crystallographic lattice. One can, easily, imagine that in such a scenario the gap along the x and the y directions might be at the same energy, i.e. degenerate. These two cases can lead to chiral superconductors of the form $p$+$ip$ or $d$+$id$, respectively. Such a scenario is schematically depicted in figures \ref{fig:3}(d)\&(e). If one finds a layered chiral superconductor of such a sort, it is likely that this chiral superconductor can be topologically non-trivial, or that it can be manipulated into a topologically non-trivial state by changing chemically the interlayer distance, and thereby the interlayer interactions. Given our current knowledge, these two cases are the most realistic scenarios for realizing intrinsic topological superconductivity.

\subsubsection*{Helical Topological Superconductors}

The analogue to the 3D topological insulator for a system with a superconducting gap is the helical superconductor \cite{Martin2012,Pientka2013,Ozfidan2016}. For this, a helical-pairing superconducting state has to be realized with a winding number. In this state, the fermions with up spins are paired in the $p$+$ip$ state, and fermions with down spins are paired in the $p$-$ip$ state. In 3D this leads to a time-reversal-invariant state, which is fully gapped. This state is very well studied theoretically and there are proposals for various systems, such as e.g. the non-centrosymmetric superconductor \ce{CePt3Si} \cite{Yanase2008,Scaffidi2014}.

The surface of a 3D topological superconductor would contain Dirac cones of Bogoliubov–de Gennes particles, i.e. 2+1D helical surface fermions. To obtain Majorana zero-modes should be possible, however, it would likely require low-dimensional nano-engineered constrictions. Alternatively, one can also envision a 2D helical superconducting state, which would also be topologically non-trivial in 3D if the interaction between these 2D sheets would be weak. Such a helical $p$-wave superconductor in 2D is depicted in figure \ref{fig:3}(f).

Neither of these helical superconducting states have so far not been observed conclusively. In the current literature, there is only limited data available on the necessary ingredients to obtain a helical superconductor. The most important prerequisite is the occurrence of a triplet superconducting state. This state has to occur in combination with either considerable Hund's coupling --- but too-large Hund's coupling, of course, leads to a magnetically ordered state, which competes with the superconducting ground state --- or within a non-centrosymmetric space group \cite{Bauer2012}. For example, the lack of inversion symmetry might transform a chiral $p$-wave superconductor into a helical one. Henceforth, with a strong 2D character and weak interactions of the helical superconducting planes, one might also be able to find -- analogue to a weak topological insulator --- in a stacked series of layers of 2D helical superconductors. This topologically non-trivial phase is time-reversal symmetry breaking.

\subsubsection*{Other Topological Superconductors}

Beyond these strong-topological, fully-gapped phases described above, there have also weak topological phases, Weyl superconductors, higher-order topological phases, and other topologically non-trivial superconducting states been proposed. For example, in the so-called higher-order topological superconductors, there one takes into account crystalline symmetries in addition to non-spatial ones.\cite{geier2018second,langbehn2017reflection,ono2020refined} These resulting higher-order topological superconductors have boundary states with multiple dimensions that are non-trivial: in two dimensions, the second-order topological superconducting phase has a gapped spectrum of propagating modes along the edge, zero-energy modes localized at the corners of the system, and a gapped bulk quasi-particle spectrum, hence Majorana hinge and corner modes are expected in such materials. 

While these phases are highly interesting and host various unique aspects of previously unexplored physics, it may from a chemists' perspective be fair to say that such materials might be considered as a more complex version of the fully-gapped, strong topological superconductors as described above. Henceforth, in this perspective -- that aims at enlightening the requirements for an intrinsic topological superconductor -- we will not further focus on the requirements necessary for realizing these more special cases of topological superconductors.

\section{How to Proof Topological Superconductivity}

Evidence for the discovery of an intrinsic topological superconductor will eventually come from a series of detailed measurements of physical properties. Proving the existence of a topological superconductor is an iterative process that requires a combination of experimental and theoretical techniques, as well as collaboration between different research groups. It is important to be cautious and thorough in the analysis to avoid drawing premature conclusions. To provide unequivocal evidence for the discovery of topological superconductivity will require edge and surface sensitive measurements. These include in-depth scanning tunneling microscopy (STM) and tunneling spectroscopy. STM allows for the analysis of the unconventional superconductivity as it can provide evidence of line nodes from quasiparticle tunneling spectroscopy, and allows for the analysis of the gap over $T_{\rm c}$ ratio, which is a great indication of unconventional pairing mechanisms, as the BCS theory predicts a fixed ratio. For a topological superconductor, one needs to prove -- using STM -- that there is a zero-biased conductance peak. This zero-biased conductance peak is governed by some specific field dependencies, including spin-orbit anisotropy and peak oscillations. Furthermore, it has to be proven that there is another Majorana on the opposite edge/surface.\cite{lee2014spin,frolov2020topological} Final evidence for a topological superconductor will be gained through the observation of surface currents within the superconducting state. These can be detected by highly specialized surface sensitive transport measurements.\cite{Schnyder2013,Sato2017}

These direct proof methods are commonly beyond the capabilities of a usual chemical physics or physical chemistry laboratory. However, there are many properties and indications that can be observed in a superconductor that may indicate proposing these measurements to collaborators, and which can be measured in a straightforward fashion. Certainly, high-quality, phase-pure samples and/or crystals -- which are characterized in-depth by crystallography -- are an essential prerequisite. This is crucial, especially for superconductivity research, as impurities or defects in the sample and/or crystal can significantly affect the experimental results and lead to incorrect conclusions. (see, e.g., reference \citenum{witteveen2021polytypism}). 

Properties that a superconductor may display that hint towards unconventional and potentially topological superconductivity include, but are not limited to: 

\textbf{(i) Linear resistivity above the critical temperature $T_{\rm c}$: } Linear resistivity is often observed in unconventional superconductors, such as high-temperature superconductors like the cuprates or the iron-based superconductors. The linear temperature dependence of resistivity in the normal state of these materials can be attributed to the presence of strong electron-electron correlations, which lead to non Fermi-liquid behavior. In conventional superconductors, the resistivity typically exhibits a different temperature dependence in the normal state. For example, in simple metals, the resistivity often follows the Bloch-Grüneisen formula, which corresponds to a $T^2$ dependence at low temperatures due to electron-phonon scattering.

\textbf{(ii) Unusual features in the specific heat:} The specific heat of a superconductor is an important thermodynamic property that can provide insights into the superconducting transition and the nature of the electron pairing. Furthermore, it is widely considered the critical proof for a superconductor being a bulk superconductor.\cite{carnicom2019importance,witteveen2021polytypism} The behavior of the specific heat below the critical temperature $T_{\rm c}$ can be qualitatively different for conventional and unconventional superconductors. Below $T_{\rm c}$, the specific heat of a superconductor can be described by the following expression:\cite{Tinkham2004}

\begin{equation}
C(T) = C_n(T) - \alpha T^3 \exp{\left(-\frac{\Delta(0)}{k_B T}\right)},
\end{equation}

where $C_n(T)$ is the specific heat in the normal state, $\alpha$ is a constant, $\Delta(0)$ is the superconducting energy gap at zero temperature, and $k_{\rm B}$ is the Boltzmann constant. The exponential term in this expression reflects the suppression of low-energy excitations due to the formation of Cooper pairs. As the temperature approaches zero ($T \rightarrow 0$), the specific heat of the superconductor decreases exponentially, eventually becoming negligible. In unconventional superconductors, the specific heat behavior below $T_{\rm c}$ can be deviating from the above behavior, reflecting the presence of competing orders or of a non-trivial gap structure. The specific heat allows us to differentiate between nodal and node-less superconductors, which might point to unconventional pairing.

\textbf{(iii) Large upper critical field $H_{\rm c2}$:} The upper critical field ($H_{\rm c2}$) can provide important information about the nature of superconductivity in the material. Unconventional superconductivity can be related to unusually small and/or large critical fields. In particular, a large upper critical field ($H_{\rm c2}$) is widely considered to be indicative of an unconventional superconductor.\cite{Hunte2008,Bristow2020,falson2020type,ma2021superconductivity} The maximal upper critical field $H_{\rm c2}$ in the weak-coupling Bardeen-Schrieffer-Cooper (BCS) theory for superconductivity is given by the paramagnetic pair breaking effect, which is commonly known as the Pauli paramagnetic limit $H_{\rm Pauli}$\cite{Tinkham2004}. This value is given by 

\begin{equation}
    \mu_0 H_{\rm Pauli} = \frac{\Delta_0}{\sqrt{2} \mu_{\rm B}} \approx 1.86{\rm [T/K]} \cdot T_{\rm c}
\end{equation}

with $\mu_{\rm B}$ being the Bohr magneton and $\Delta_0$ being the superconducting gap. In conventional superconductors, the upper critical field is limited by the Pauli paramagnetic limit, which arises due to the competition between the superconducting pairing and the Zeeman energy of the electrons in the magnetic field. The Pauli paramagnetic limit can be higher for strong-coupled superconductors \cite{clogston1962upper,orlando1979critical}. However, some superconductors may exceed this limit significantly, such as those with spin-triplet pairing or highly anisotropic gap structures. 

\textbf{(iv) Unusual temperature dependence of the upper critical field:} The upper critical field $H_{\rm c2}$ separates the mixed state (or vortex state) from the normal state. The temperature dependence of $H_{\rm c2}(T)$ can be described by the Ginzburg-Landau theory or the Werthamer-Helfand-Hohenberg (WHH) theory for conventional superconductors.\cite{werthamer1966temperature} For unconventional superconductors, the temperature dependence of $H_{\rm c2}(T)$ may deviate from these models due to the non-trivial gap structure, strong electron correlations, and/or unconventional pairing mechanisms.

\textbf{(v) A large value for the normalized specific heat jump:} The BCS weak-coupling value for the normalized specific heat jump is $\Delta C/\gamma T_{\rm c}$ = 1.43. If this value is much larger than either $\gamma$ is very large or the specific heat jump $\Delta C$ is very small. Both cases are unusual and hint towards unconventional superconducting properties. In the case of a large, $\gamma$ it is worthwhile to analyze, whether the material is a heavy fermion compound by determining the Kadowaki-Woods ratio and the Wilson ratio.\cite{kadowaki_universal_1986,coleman_heavy_2007,von_rohr_superconductivity_2014,lefevre2022heavy} This may indicate strong electron-electron correlations.

\textbf{(vi) Low carrier-density superconductors:} While a very high density of states can hint towards unconventional superconductivity, so can also an uncommonly low-density of states of a superconductor. A low carrier density is associated with unconventional superconductivity because it is an indication that the superconducting mechanism may not be governed solely by phonon-mediated electron pairing.\cite{caviglia2008electric,arpino2014evidence} A low carrier-density is reflected in a low $\gamma$ in the specific heat as $\gamma \propto D(E_{\rm F})$, or directly by Hall effect measurements.

\textbf{(vii) Competing orders in a material or in the phase diagram:} This is relevant, when other orders than superconductivity are observed in a material, e.g., local or long-range magnetic order, charge order, or orbital order. Unconventional superconductors are commonly found to exhibit competing orders, which are believed to share a common driving force, such as strong electronic correlations or proximity to a quantum critical point. 

\textbf{(viii) Flat bands close to the Fermi-level:} When band-structure calculations reveal flat bands and/or van-Hove singularities at or close to the Fermi-level. Flat bands are considered an indication of strong electron-electron correlations due to their high density of states, reduced kinetic energy, potential localization, and the emergence of competing or cooperative interactions. Hence, flat bands and/or van-Hove singularities are widely believed to be hinting towards strong electron-electron interactions.

\textbf{(ix) More than one superconducting transition:}  It is considered to be likely that below the transition temperature for example a $d$-wave symmetric gap appears first, before a transition to a chiral $d$+$id$ symmetric gap occurs at lower temperatures (see, e.g., reference \cite{Kiesel2013}). Such a transition should be visible in the magnetization and certainly in the specific heat, as it is related to an entropy change.

\textbf{(x) Square, hexagonal, triangular, or kagome layers:} If superconductivity with any of the features described above are observed on square, hexagonal, triangular, or kagome layer, then it is clearly worth to take a closer look at the respective material. Due to symmetry reasons, it may be more likely to find a degenerate gap on these crystallographic structures (see, discussion above).

Further, evidence for an unconventional superconducting state can be found by means of the microscopic techniques -- which are nowadays widely available and accessible also to chemists. These are, specifically, muon-spin-rotation ($\mu$SR) and nuclear magnetic resonance measurements (NMR). $\mu$SR allows for the measurement of the London penetration depth and is a highly sensitive technique to observe time-reversal-symmetry breaking. Furthermore, it allows an analysis of potential unconventional superconducting properties by the measurement of the superfluid density $\rho(T)$, e.g. in the "Uemura plot". NMR in superconductors allows measuring local magnetic fields, spin relaxation rates, and Knight shifts to reveal insights into the electronic behavior, magnetic properties, and pairing mechanisms of these materials.

\section{Current Candidates for Topological Superconductors}

There are a handful of candidates to date, considered as potential intrinsic topological superconductors. Several compounds display signs of unconventional, potentially topological superconductivity. The results from different experimental methods probing their potentially topological superconducting properties are, however, contradictory, and as a consequence, it should be far to say that a material that convincingly displays intrinsic topological superconductivity -- and the resulting Majorana zero modes -- has so far not been discovered. In the following, we will discuss a selection of the potential topological superconductors that are discussed in the most recent literature.

\subsection*{Perovskite Oxides}

\ce{Sr2RuO4} has been considered a prototypical unconventional superconductor.\cite{mackenzie2017even} It was long believed to be the only solid-state analogue to the superfluid \ce{^3He} A-phase.\cite{rice1995,luke1998time} This insight made it into numerous textbooks. However, in a recent revisit of NMR experiments the Knight shift -- a measure of spin polarizability -- was found to decrease at temperatures below the critical temperature, consistent with a drop in spin polarization in the superconducting state.\cite{pustogow2019constraints} Hence, the most prevalent theoretical interpretation of the order parameter as a chiral $p$-wave state were found to be insufficient.\cite{kivelson2020proposal} Such a $p+ip$ superconducting order parameter was long believed to be close to a topological superconducting state. Now, even though such an order parameter can likely be excluded \ce{Sr2RuO4}, nevertheless, remains an extremely clean layered superconductor that emerges from a strongly correlated Fermi liquid. This makes this material a promising candidate to find topological superconductivity. One way to achieve this is considered via an $d + id$ order parameter, which seems to be in agreement with many of the experimental observations at hand. To summarize all the current work on \ce{Sr2RuO4} would be beyond the scope of this review and the interested reader is referred to references \citenum{mackenzie2017even,kivelson2020proposal}. 

Similarly, with the same and/or related structures, the cuprates might be promising platforms to realize topological superconductivity if the right symmetries and electronic structures are met.\cite{lu2014underdoped} Of especial interest in this regard has been the proposal of realizing topological superconductor in a twisted bilayer of cuprate sheets.\cite{can2021high} This proposal itself is certainly not an intrinsic topological superconductor, however, such a crystal structure with twisted layers of metal oxide planes, may not seem inaccessible -- given the right synthesis conditions -- also as a bulk crystal.  

\subsection*{Transition Metal Dichalcogenides}

A promising class for the discovery of an intrinsic topological superconductor can be found in the family of transition metal dichalcogenides (TMDs). TMDs have attracted a lot of attention in recent years due to their unique and rich electronic properties.\cite{devarakonda2020clean,coleman2011two,Guguchia2017} TMDs are known to display charge ordering, various types of long-range magnetic ordering, Mott-insulating states, as well as superconductivity.\cite{law20171t,von2019unconventional,guguchia2018magnetism,nicholson2021uniaxial,costanzo2016gate} The (co-)existence of various complex electronic and magnetic states in these materials, coupled with a great tunability of these interactions, due to the heavily layered structure, make them ideal candidates for the realization of most unconventional electronic and magnetic phases.\cite{hsu2017topological}

Among the TMDs that are currently considered as possible topological superconductors are among others 4H$_{\rm b}$-\ce{TaS2} and T$_{\rm d}$-\ce{MoTe2}. 4H$_{\rm b}$-\ce{TaS2} crystallizes in the hexagonal space group $P63/mmc$.\cite{di1973preparation} Its unit cell consists of alternating layers of 1H-\ce{TaS2} and 1T-\ce{TaS2}. Hence, 4H$_{\rm b}$-\ce{TaSe2} can therefore be considered as a naturally occurring heterostructure between a 2D superconductor (1H layers) and a doped Mott insulator, proposed to be a gapless spin liquid (1T layers). 4H$_{\rm b}$-\ce{TaSe2} is a superconductor with a  critical temperature of $T_{\rm c} \approx$ 2.7 K. Recently, it has been found that there are signs of time-reversal symmetry breaking at the critical temperature.\cite{Ribak2019} Given the hexagonal symmetry and the Fermi surface topology, these findings suggest that 4H$_{\rm b}$-\ce{TaS2} might be a highly layered chiral superconductor, which might allow for the observation of a topological superconductor in this system if weak interlayer coupling is present.

T$_{\rm d}$-\ce{MoTe2} is the low-temperature structure of 1T'-\ce{MoTe2}. It is a Weyl semimetal, i.e., it exhibits a topological phase transition, resulting in the formation of Weyl points in the electronic structure.\cite{tamai2016fermi,soluyanov2015type,guguchia2020pressure} These points act as sources or sinks of Berry curvature, which is a measure of the geometric phase acquired by the electrons as they move through the material. The presence of Weyl points leads to the emergence of chiral surface states, i.e., Fermi arcs. To date, there are only few Weyl superconductors known. T$_{\rm d}$-\ce{MoTe2} is clearly the most prominent among them, especially as indications for topological superconductivity were found. Several unusual superconducting properties were observed in this material, including robust edge supercurrents, two-gap superconducting order parameter, strong dependence of the critical temperature on sample quality, and the enhancement of the critical temperature under pressure and in the monolayer limit.\cite{wang2020evidence,qi2016superconductivity,Guguchia2017,rhodes2021enhanced}

\subsection*{Intercalated \ce{Bi2Se3}}

Bismuth selenide had been one of the first topological insulators that had been discovered. Topologically protected Dirac cone surface states have been predicted and observed in this material.\cite{zhang2009topological,xia2009observation,hasan2010colloquium} Shortly after this discovery, it was found that intercalated \ce{Bi2Se3} becomes superconducting.\cite{hor2010superconductivity,sasaki2011topological} Superconductivity in \ce{A_xBi2Se3} has been observed for the dopants A = Cu, Nb, and Sr. All of these superconductors display critical temperatures up to approximately 4 K. They all exhibit superconductivity with a remarkably low carrier density at the Fermi level. These compounds are intercalated versions of, \ce{Bi2Se3} consisting of triangular-lattice layers of Bi and Se. Where the intercalant is occupying a position in the van der Waals gap.

The realization of a broken-symmetry state in proximity to topologically protected spin channels was and still is interesting, and might be a manifestation of the Fu-Kane proposal to realize topological superconductivity by a proximity effect. But beyond that \textit{Liang Fu} and \textit{Erez Berg} showed, in fact, at the example of \ce{Cu_xBi2Se3} that one might obtain an odd-parity topological superconductors.\cite{Fu2010} In their work, the authors showed that in the case of odd-parity a topological superconductor can be realized, depending on the number of Fermi-pockets. A prerequisite for this would be $p$-wave superconductivity and a peculiar Dirac band structure, as proposed for \ce{Cu_xBi2Se3}.

There are experimental indications for unconventional, potentially topological superconductivity in these materials. However, the chemistry of this material seems to pose one of the grand challenges for clear evidence for an unconventional or even topological superconducting state. The synthesis of superconducting \ce{A_{x}Bi2Se3} has been found to be highly difficult, and the availability of materials with reproducible properties cannot be guaranteed. The best sample qualities were obtained by employing an electrochemical intercalation methodologies.\cite{Kriener2011} A better materials platform related to these superconductors would be highly desirable.

\subsection*{Superconducting Topological Semimetals}

Topological semimetals are a particularly promising way to go when it comes to topological materials that also exhibit superconductivity. In contrary to e.g., topological insulators as the challenge for \ce{A_xBi2Se3} (see above), topological semimetal have a higher electronic density of states at the Fermi level, making it more feasible for them to exhibit superconductivity, and unnecessary to first dope them into a metallic regime. A fairly wide range of topological semimetals have been investigated and characterized, with the aim for discovering unconventional superconducting properties, and eventually topological superconductivity. Examples of these materials include the aforementioned Weyl semimetal T$_{\rm d}$-\ce{MoTe2}\cite{qi2016superconductivity}, the Dirac semimetal \ce{PdTe2} which exhibits both type-I as well as type-II superconductivity\cite{clark2018fermiology}, or the nodal-line semimetal NaAlSi, which was found to be a two-gap superconductor with a critical temperature of $T_{\rm c} \approx$ 9 K \cite{Muechler2019}. Of great interest has also been the Dirac semimetal 2M-\ce{WS2}, which consists of 1T'-\ce{WS2} monolayers, with a critical temperature of $T_{\rm c}$ = 8.8 K\cite{li2021observation,Guguchia2019}, and the half-heusler superconductors REPtBi and REPdBi, which show different topological bands, as well as coexistent superconductivity and long-range magnetic order\cite{kim2018beyond,tafti2013superconductivity}. Many of these superconducting topological semimetals display unconventional superconducting properties, which makes them promising candidates for topological superconductivity.

\subsection*{Superconductors with Honeycomb Layers}

The honeycomb lattice has attracted immense interest in the material science community over the years.\cite{takagi2019concept,weller2005superconductivity,nagamatsu2001superconductivity,felser1999electronic} Materials with such layered structures can host a multitude of intriguing physical phenomena, for example they have topological band structures, some of them are quantum spin liquid candidates, and there are well-known superconductors with honeycomb structures.\cite{Banerjee2017,do2017majorana,zhang2011topological} Much of the physics that governs this type of layered structures has been worked out and identified for graphene, but may also be applied to other related materials. 

The honeycomb layered structure has also been identified as a promising platform for hosting unconventional and topological superconductivity.\cite{xu2018topological,kiesel2012competing,black2014chiral} It has been shown theoretically that antisymmetric spin–orbit coupling plays an important role in locally non-centrosymmetric superconductors.\cite{fischer2022superconductivity,sigrist2014superconductors} These non-centrosymmetric superconductors are made up of non-centrosymmetric sublattices but retain the global space inversion center. A widely discussed example for this has been SrPtAs, which crystallizes in an ordered variant of the famous \ce{AlB2}-type structure with honeycomb layers consisting of Pt and As.\cite{nishikubo2011superconductivity} SrPtAs was found to be a superconductor with a critical temperature of $T_{\rm c}$ = 2.4 K. Furthermore, it was found that at the superconducting state breaks time-reversal symmetry.\cite{biswas2013evidence} Hence, experimental data and theoretical work suggests that SrPtAs might be a chiral $d$-wave superconductor. Hence, if the coupling between the PtAs layers is weak enough, this compound might be a topological superconductor.\cite{fischer2014chiral} These \ce{AlB2}-type ternary compounds are a versatile, promising playground for the discovery of unconventional, potentially topological superconductivity.\cite{evans2009structural,walicka2021two} This large family of materials are highly tunable by synthesis conditions and precise chemical composition, e.g. slightly off-stoichiometric SrPtAs is not crystallizing in a non-symmorphic space group anymore, or some of these materials might display buckled hexagonal layers.

\subsection*{Delafossite-Type Superconductors}

Delafossite-type materials are compounds of the type \ce{ABX2}, where A and B are different metallic elements, and X is a chalcogen.\cite{shannon1971chemistry} The crystal structure of these materials consists of an alternate stacking of triangular A lattices and planes of edge-sharing \ce{BX6} octahedra along the $c$-axis. These two different layer types are connected via the chalcogen in a so-called dumbbell position. There are two possible stacking scenarios: rhombohedral with $R\bar{3}m$ space-group symmetry and hexagonal with $P63/mmc$ symmetry. 

One of the most prominently investigated delafossite is sodium cobaltate \ce{Na_xCoO2}, which can host Mott physics, complex magnetism, charge- and spin-ordering, and even superconductivity when intercalated with water. \ce{Na_x(H2O)_yCoO2} is a remarkable superconductor, discovered serendipitously.\cite{foo2004charge,takada2003superconductivity} Superconductivity with a critical temperature as high as $T_{\rm c}$ = 4.5 K occurs in this material in the low Na content region and upon incorporation of water into its structure.\cite{schaak2003superconductivity} The vicinity of strongly correlated properties and frustrated magnetism, as well as this occurring on a triangular lattice, make this material, and related materials, a highly promising platform for unconventional superconductivity.\cite{lechermann2021basic,hicks2012quantum} Henceforth, a chiral $d$-wave superconducting state was proposed for this compound, which eventually might also be topologically non-trivial.\cite{Kiesel2013} The realization of high-quality reproducible samples of \ce{Na_x(H2O)_yCoO2}, specifically, has been a challenge for this material and makes a clear identification and characterization of its properties challenging to date. The discovery of another superconductor in this family would be fascinating.

\subsection*{Uranium-Based Heavy-Fermion Superconductors}

Spin-triplet superconductors are strong candidates for topological superconductors. They can either be topological as a chiral $p+ip$-type superconductor with sheets that are weakly interlinked, or in the form of a helical superconductor. The heavy fermion superconductors \ce{UPt3}, \ce{UGe2}, \ce{URhGe}, \ce{UCoGe}, and \ce{UTe2} are currently strong candidates for $p$-wave superconductivity.\cite{ran2019nearly,huxley2001uge,tsutsumi2012spin,aoki2001coexistence,huy2008unusual} They all exhibit low superconducting critical temperatures, which makes the probing of the superconducting state certainly challenging.

Superconductivity in \ce{UTe2} was discovered in 2018 with a transition temperature of $T_{\rm c}$ = 1.6 K. It has the highest critical temperature among these materials, which makes it a highly interesting platform to be studied. While there remain discussions about the superconducting order parameter of \ce{UTe2}, this material is widely considered to be a $p$-wave superconductor. Evidence for this includes that the upper critical field $H_{\rm c2}$ exceeds by an order of magnitude the Pauli paramagnetic limit and by the absence of a Knight shift in the NMR data at the critical temperature.\cite{ran2019nearly,ran2019extreme,nakamine2019superconducting}
Claims of topological superconductivity are related to the detection of signatures of chiral surface excitations, as well as the detection of chiral state in London penetration depth measurements.\cite{jiao2020chiral} Overall these uranium-based heavy-fermion superconductors are certainly a promising platform for the realization and/or observation of a topological superconductor, while higher temperatures for their critical temperatures would be certainly advantageous for characterizing the properties in detail.

\subsection*{Kagome Superconductors}

Metallic magnets and superconductors with a kagome lattice have attracted considerable interest recently. The experimental field stated in 2011 with the ferromagnet \ce{Fe3Sn2}, though theoretical work is dating back earlier.\cite{ye2018massive,mazin2014theoretical,kiesel2012sublattice}

Kagome structures have long been a study focus for the search of frustrated magnetic systems and eventually for the realization of a quantum spin liquid.\cite{zhou2017quantum} The kagome structure is characterized by a hexagonal lattice of triangles, with each atom having four neighbors in the same plane. More recently, it was realized that the unique combination of hexagonal symmetry and band folding can also result in a range of intriguing properties, such as the occurrence of massless and massive Dirac fermions, with a characteristic band structure consisting of linearly crossing bands, as well as a flat band that might be a key ingredient for correlated physics.\cite{yin2020quantum,guo2022switchable,mielke2022time} Especially the discovery of superconductivity in the kagome compounds \ce{AV3Sb5} (A = K, Rb, Cs).\cite{neupert2022charge} These compounds have not only been found to exhibit unconventional chiral charge order, but also display unconventional superconducting properties, including indications of a nodal energy gap and time-reversal symmetry breaking.\cite{wu2021nature,guguchia2023tunable} It remains possible that these compounds are conventional $s$-wave superconductors, which would make them topologically trivial. However, the kagome lattice as a platform to construct a topological superconductor remains highly promising. A kagome lattice necessarily leads to a flat band, i.e. high correlations, and a $p$-wave or $d$-wave pairing would be degenerate, i.e. result in a chiral superconductor. Hence, finding new kagome superconductors is highly desirable, as they hold excellent ingredients to host a topological superconductor.

\section{Key Ingredients for the Discovery of an Intrinsic Topological Superconductor}

\begin{figure}
	\centering
	\includegraphics[width=0.8\textwidth]{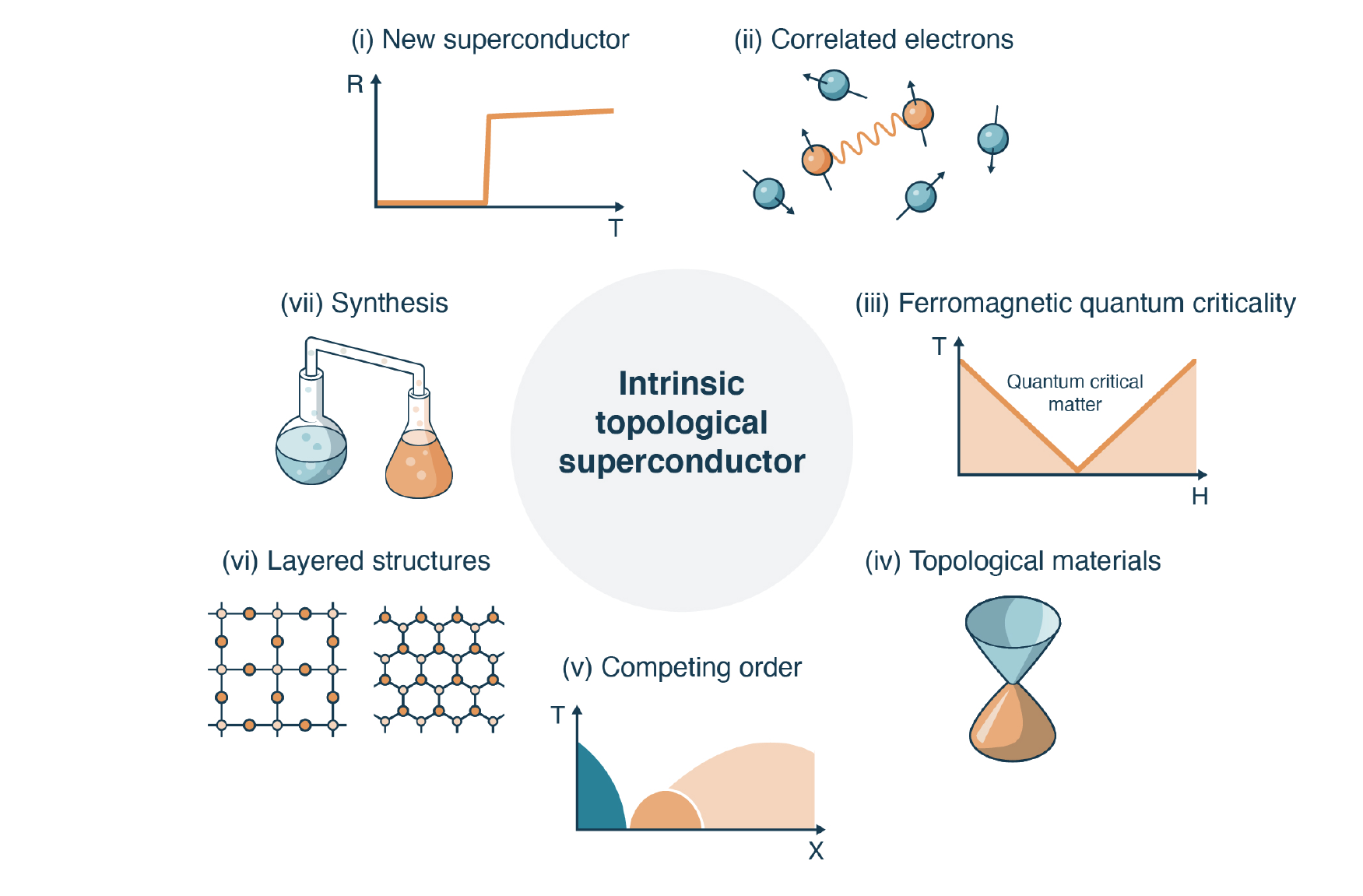}
	\captionof{figure}{Schematic representation of the key ingredients to discover a topological superconductor.}
	\label{fig:4}
\end{figure}

The materials mentioned in this perspective display unusual superconducting properties, and there are many more unusual superconducting materials out there that have not been discussed here. None of these materials have, however, been so far proven to be an intrinsic topological superconductor. Hence, the quest to find such a material remains a topic of great interest. It is my personal opinion, that solid-state chemists will play an essential role in order to realize such a discovery. Chemists, with their extensive knowledge of various materials classes, structural correlations, and techniques for compositional tuning and defect engineering, have the tools and the knowledge to significantly advance the field of topological superconductivity. In the following, I will summarize my ideas for finding such a material that are inspired by advancements in theoretical physics. These design principles, which are based on a combination of heuristic chemical and physical requirements, might be the ingredients for the discovery of an intrinsic topological superconductor. These design principles are summarized in figure \ref{fig:4} as the following:

(i) \textbf{New superconductors}: It might seem obvious, but having new superconductors is an essential key ingredient to the discovery of a topological superconductor. There are, however, only a few new superconducting materials discovered every year, and only some of them reveal unexpected superconducting properties. Especially, new unconventional superconductors that display a fully-gapped superconducting state, may very well be related to a topological superconducting state (see discussion in chapter 5).

(ii) \textbf{Correlated electrons}: A strongly correlated Fermi-liquid in the normal state of the material is likely an excellent starting point to discovery an unconventional superconductor. Phonon-mediated superconductivity is most easily leading to an $s$-wave superconducting gap, hence highly correlated electrons are a more promising route to discover topological superconductivity. It is, for example, well established that the energy scales of the Coulomb repulsion, Hund's exchange, crystal field splitting, and the orbital bandwidth in transition metal oxides can support the realization of a strongly correlated state, i.e., a Mott insulating state. Furthermore, rare-earth containing compounds have been found to display a wide-variety of correlated physics, e.g., many intermetallic compounds containing Ce or Yb are known heavy-fermion materials. Hence, such materials and/or related materials will be favorable starting points for the search for an intrinsic topological superconductor.

(iii) \textbf{Ferromagnetic quantum criticality}: Superconductivity mediated by spin fluctuations in weak and nearly ferromagnetic metals are considered a likely place for the discovery of $p$-wave superconductivity. This can be realized by the occurrence of superconductivity close to a or at a ferromagnetic quantum critical point. Hence, looking for superconductors close to a ferromagnetic instability -- preferably with a large Hund's coupling -- might be a promising route.

(iv) \textbf{Topological materials}: In order to look for topological superconductivity, it has been popular recently to look in materials with topologically non-trivial electronic structures. The general idea being that the restrictions that the electrons in the normal state experience, e.g., in polarized spin channels or by a large spin-orbit coupling, might elevate an unconventional pairing mechanism. Furthermore, finding superconductivity with materials that have a low carrier density -- i.e., they are close to being a semiconductor, which many topological materials are -- is a promising approach for the discovery of unconventional superconductivity. Both ingredients may potentially lead to topological superconductivity. While this may be a feasible approach, it also has to be noted that having a topological material is clearly not a necessary ingredient per se for realizing a topological superconductor. Also, other low-carrier materials might in this regard be interesting, such as e.g., organic superconductor, where there are likely many superconductors to be discovered yet.

(v) \textbf{Competing Order}: To have several competing orders has been found to be a major ingredient in the manifestation of unconventional superconductivity. Whether these orders are of competing, coexisting, or even enhancing nature has been a cause for major debates. What can be stated clearly is that these different orders appear experimentally in most (if not all) cases of unconventional superconductivity. For example, in the phase diagram of the prototypical cuprate superconductor \ce{La_{2-x}Sr_xCuO4} not only the unconventional $d$-wave superconductivity has been found, but also antiferromagnetic order, charge order, and the (still mysterious) pseudogap phase. Especially, charge order, charge density waves, and superconductivity follow very similar physical laws so that it remains unclear when one should be favored over the other. Although it is not necessarily required, examining materials with competing orders in proximity to superconductivity may be an effective starting point for synthetic chemists seeking to discover an unconventional, potentially topological superconductor.

(vi) \textbf{Layered structures}: Finding an intrinsic topological superconductor in a layered material with a highly anisotropic electronic structure seems most feasible, for arguments discussed above. Especially looking for hexagonal, triangular, or kagome lattices with a $d$-wave superconducting gap in which a $d+id$ wave results from the (near) degeneracy as a result of the lattice symmetry is an auspicious approach. Alternatively, a $p$-wave symmetry on a hexagonal or square lattice -- based on the same type of symmetry argument -- would be a promising approach as well. Overall, the crystal structure and the crystal symmetries are essential ingredients for the discovery of a topological superconductor, especially also global and local non-centrosymmetric structures should be carefully examined.

(vii) \textbf{Synthesis}: What kind of chemistry should we all be looking for to discover a topological superconductor? For one, there have been very exciting developments recently in the synthesis of metastable phases of quantum materials.\cite{mcqueen2021future,ma2021synthetic} Clearly, materials that require advanced synthesis methodologies involving air sensitive reagents, low-temperature reaction conditions, i.e., soft-chemical synthesis, metathesis reactions, and syntheses requiring high-pressures are not only pathways to discover new compounds, but rather also new avenues to unprecedented physics, such as a topological superconductor. Furthermore, it is apparent that quaternary and quinternary phases are to be more extensively investigated, yet. Especially mixed-anionic materials with highly tunable electronic properties are a versatile playground that needs to be explored. My personal view is that in comparison to the oxides only few nitrides, chlorides, fluorides, and bromides have been investigated. Leaving ample space for unprecedented discoveries, such as a topological superconductor. 

Hopefully, this perspective has revealed that the topic to discover is a vibrant, exciting field of research. The discovery of such a material has far-reaching implications, as there are many theoretical predictions that can be explored, tested, and even put to use, as there is an already whole research field dedicated to the realization of fault-tolerant quantum computing based on Majorana zero modes that would occur in the vortex core of such a material. I hope I could point out that this field in essence requires synthetic chemists, as discovery as well the synthetic realization of these materials is not straight-forward, and you are more than welcome to join the quest for such a material -- and discover many exciting, new materials and properties on the way.

\section*{Acknowledgements}
The author thanks Ronny Thomale, Louk Rademaker, and Mark H. Fischer for insightful discussions. This  work  was  supported by the Swiss National Science Foundation under Grant No. PCEFP2\_194183.

\bibliography{achemso-demo}

\end{document}